\documentstyle[12pt]{article} 
\begin{document}
\title{Determination of Finite Size Effects in Lattice Models from 
the Local Height Difference Distribution }  
\author{S.V. Ghaisas\\ 
 Department of Electronic Science, University of Pune, Pune 411007,
India
}

                  
\maketitle 

 Growth of interfaces during vapor deposition is analyzed on a 
discrete lattice.  It leads to finding distribution  
of local heights, measurable for any lattice model. Invariance in the 
change of this distribution in time is used to determine the finite 
size effects in various models  The analysis is applied to the discrete linear growth equation
and  Kardar-Parisi-Zhang (KPZ) equation.
 A new model is devised that shows early convergence to the KPZ dynamics. 
 Various known conservative and non conservative models 
are tested on a one 
dimensional substrate by comparing the growth results with the exact KPZ 
and linear growth equation results. 
The comparison helps in establishing the condition that helps in determining 
the presence of finite size effect for the given model. The new model is used 
in 2+1 dimensions to predict close to the true value 
of roughness constant for KPZ equation.


{pacs}{60., 68.55-a,82.20.Fd}   
\vfill
\pagebreak                          
Growth on a lattice from vapor can be represented in primarily two ways. It can be 
modeled as a lattice model where the atomic interactions are replaced by 
simple growth rules\cite{bar,kr2} , then obtain a growth equation based on various 
symmetries of the problem under consideration\cite{bar}. 
Other way is to construct the growth terms from the 
given growth rules for a lattice model at the coerce-grained time and length 
scales\cite{sv1}. The KPZ equation was introduced to include lateral growth 
in growth equation \cite{kpz}. It has attracted a lot of attention 
in the field of growth. There are many lattice models and numerical 
solutions claiming to belong to the same universality as KPZ 
equation\cite{kk,ml,mpp,ggr,bd,other,cast}. 
In 1+1 dimensions,  exponents  can 
be exactly obtained\cite{bar}. However, in higher dimensions exact values are 
not obtained. Various lattice models and numerical solutions predict a range 
of values due to the finite size corrections. In the following we develop a method to 
determine the existence of finite size effect in a model. A model that 
converges to its representative universality can be identified and hence correct exponents 
can be determined.

 A linear equation representing 
interface motion  normal 
to the surface can be obtained in the frame of reference moving 
with the interface velocity by considering inter-planer hopping of ad atoms
on the interface with a bias for downward or in-plane hopping toward step edge
 \cite{sv1}.   
It has the form 
\begin{equation}
    \frac{\partial h}{\partial t}=\nu_{0}\nabla^{2}h+\eta  
\end{equation} 
where, $\nu_{0}$ explicitly depends upon $F$, and $\eta$ is the noise 
due to the randomness in the deposition flux. It has the correlation given 
by $<\eta({\bf x},t)\eta({\bf x'},t')>=2D\delta({\bf x-x'})\delta(t-t')$. The 
angular brackets denote the ensemble average of the contents. Eq. (1) is known as 
Edward-Wilkinson (EW) equation \cite{ew}. The lowest ordered non-linear 
correction to EW equation was introduced by Kardar , Parisi , and Zhang 
\cite{kpz}. The resulting equation, 
\begin{equation}
    \frac{\partial h}{\partial t}=\nu_{0}\nabla^{2}h+ \lambda(\nabla h)^{2}
                     +  \eta  
\end{equation} 
 is known as KPZ equation. This is a non-conservative equation.  

The steady state growth is characterized by roughness exponent $\alpha$ and 
$z$ , determining the evolution correlations in time. One can measure $\alpha$ from the 
height-height (h-h) correlations, 
 
\begin{eqnarray}
G(x,t)&=&\frac{1}{N}\sum_{x'}(h(x+x',t)-h(x',t))^{2}\nonumber \\
&=&x^{2\alpha}f\left(\frac{x}{\xi(t)}\right)
\end{eqnarray}

where, correlation length $\xi(t)\sim t^{1/z}$. In the limit $x\rightarrow 
0$, $f\rightarrow 1$. Thus for a large $\xi(t)$, the plot of $G(x,t)$ vs.$x$ 
on the log scale must be a straight line for small $x$ on any scalable surface. 
Hence any lattice model should comply with this requirement for large enough 
length and time scales. Absence of straight region over large enough length 
and time scales for a lattice model indicates that the corresponding surface 
is not scalable and hence such a model cannot exactly follow the growth 
equation that it is supposed to represent. We elucidate this point 
in the case of models believed to represent 
EW and KPZ equations.  
Time exponent $\beta$, where $z=\alpha/\beta$ can be 
obtained by measuring the width over a substrate of length $L$ as, 
$w_{2}=\frac{1}{N}\sum_{x}(h(x,t)-\bar h)^{2} 
=L^{2\alpha}g\left(\frac{L}{\xi(t)}\right)$,
It can be shown that \cite{bar} for small times $w_{2}\sim t^{2\beta}$.

 We first analyze the $G(x,t)$ for some of the models representing KPZ and EW type growth
. Finite size effects enter due to 
both , the substrate size $L$ and the cutoff length $a$. The rules in a lattice 
model are sensitive to both these lengths. This results in to deviation of the 
growth dynamics from that of the growth equation, that the model is representing. 
Signature of this deviation is obtained in the non linear behavior of $G(x,t)$
in the limit $x<<\xi$ where $\xi$ is correlation length, 
from a log-log plot of $G(x,t)$ Vs. $x$ . We have measured this deviation by fitting 
straight lines on log-log plot of $G(x,t)$ Vs. $x$ for every interval of $\Delta x
=$10. From these straight line fits $y$=$mx+c_{0}$, we obtain $c_{0}$ as a function 
of average $x$. In the absence of a curvature in the log-log plot of $G(x,t)$ Vs. $x$
, one must obtain $c_{0}$ to be independent of average $x$.  

The models are briefly described below. In most of the known models\cite{bar}, finite size 
effects are present. We have chosen the model introduced in reference \cite{kk} along with 
a new model. This new model, we believe, converges to the KPZ dynamics during early growth 
as will be seen from the results ahead.  

a) KK model \cite{kk}: In this model growth proceeds by selecting a site randomly 
(this is the first step in all the models described here.). A particle is 
accommodated at the site if the absolute height difference between the selected 
site after deposition and for each of the nearest neighbors is less than or 
equal to a number $N$.

b) SC model:  We introduce another SOS model which provides limited tunability with respect to 
the spread in the distribution. This helps in identifying exponent values close to 
the true values in 2+1 dimensions. The deposition rules for the model are as follows.  
In 1+1 dimensions the deposited atom is accommodated 
if both its neighbors have at least same height as the deposited one. Otherwise, 
largest of the step differences at the site ,$s_{d}$, is obtained and accommodation is 
allowed according to the probability factor $e^{-s_{d}^2/(2\sigma ^{2})}$. Here 
$\sigma$ can be varied as a tunable parameter. In 2+1 dimensions the deposited 
atom is accommodated if three or more neighbors have at least same height as its own. 
For other depositions the accommodation is decided from the largest of the four steps 
around the site using above exponential probability factor. Details of this model are described 
elsewhere \cite{sv3}  

c) NN1 model \cite{sv1} : This is a  conservative SOS model. A particle after deposition is allowed 
to relax by hopping to a nearest neighbor site if it can lower its height. The hop 
is not allowed if the height of one or more of  its nearest neighbors is equal or larger. 

d) HM model \cite{kd1}: This conservative model is based on the models proposed in ref \cite{kd1}. 
Here, in a growth equation that involves terms of the form $\nabla^{2}f(x)$, the 
growth proceeds by allowing the particle to hop to the nearest site that has 
minimum value for $f(x)$. Thus, $f(x)$ is like a potential. For $\nabla^{2}h$, 
$f(x)=h(x)$. For, $\nabla^{4}h$, $f(x)=-\nabla^{2}h$.  
 
Models (a),(b) are assumed to belong to KPZ universality and models (d) and (e) 
to EW universality.

 Fig. 1 shows the log-log plots 
of $G(x,t)$ Vs. $x$ for various models. The substrate lengths 
and the number of MLs is large enough to 
provide saturation of $G(x,t)$ around $x=1000$.  
$c_{0}$ is measured between $x=$10 to 100. Fig. 2 shows plot of $c_{0}$ for various models 
as a function of average $x$. As can be seen, KK and NN1 models do not show straight 
line behaviour in this length scales. SC and HM models are straight lines within 
the statistical error bars. Deviation from straight line behavior is the indication 
of non scalable dynamics of the growth due to finite size effects in these models. 
Clearly, exponents derived from models like HM or SC are reliable in the respective 
universality classes in 1+1 dimensions since , from the straight 
line behavior, these models follow the 
dynamics of the growth equation that they represent. 

 This method requires measurement of $G(x,t)$ over large substrate length and 
large times. In higher dimensions it is increasingly difficult to perform such 
measurements. In order to facilitate the determination of finite size effects in 
higher dimensions, we introduce another measurement based upon the time invariance of 
distribution of height fluctuations.   

 Consider a one dimensional scalable lattice with a lattice constant $a$.  
 We define step at site $i$ as
\begin{equation}
\delta x_{i}=h_{i}-h_{i+1}
\end{equation}
The local slope is then $-\delta x_{i}/a$.
 Consider linear growth equation Eq. (1) in 1+1 dimensions.
 $ \frac{\partial h}{\partial t}\rightarrow (h_{i}(t+\Delta t)-h_{i}(t))/\Delta t$
on discretization. The r.h.s. is $(\delta x_{i-1}(t)-\delta x_{i}(t))/a+\eta_{i}(t)$.
This relation predicts the value of $h$ at $(t+\Delta t)$ dependent on the
discrete differences in steps at $t$. Let $\Delta h_{i}=h_{i}(t+\Delta t)-h_{i}(t)$.
From the nature of the equation $<\Delta h_{i}>=0$. Further, function $h_{i}(t)=
H_{i}(t)-\bar H(t)$, where $H_{i}(t)$ and $\bar H(t)$ are values of height and
average height measured from the substrate. 
  On the r.h.s. of the discrete linear equation,
$<\delta x_{i-1}-\delta x_{i}>=0$ 
and $<\eta_{i}>=0$. Thus the differential term
and the noise term can be averaged to zero independently. In fact this is true for any
conservative differential term. This observation is related to the fact that noise
does not couple to conservative differential terms except with $q=0$ mode \cite{nocop}.
For KPZ equation r.h.s. is proportional to $(\delta x_{i-1}-\delta x_{i}+\delta x_{i}^{2}
+\eta _{i})$. Between the terms $(\delta x_{i-1}-\delta x_{i})$ and $(\delta x_{i}^{2})$
, the latter term is $b^{\alpha}$ times stronger where $b$ is scaling parameter.
 Hence for large $b$,  $\Delta h_{i}$
will be determined by $(\delta x_{i}^{2})$. We will therefore consider only non linear 
term contributing to  $\Delta h_{i}$ in KPZ equation.  
 The difference between the conservative terms and
the KPZ term is that, on the r.h.s. of the equation, $<(\delta x_{i})^{2}>\ne 0$.
Since average over l.h.s. is zero, we must have $<(\delta x_{i})^{2}+\eta_{i}>=0$. For a
conservative term, the average over the term and the noise are independently zero.
Hence, in the case of KPZ equation, the KPZ term couples with noise \cite{kr2}.
In order that the average on r.h.s. be zero, evaporation or vacancy addition is
associated with the growth process .  

The distribution of $\Delta h_{i}$ in the case of KPZ equation is determined by 
$(\delta x_{i})^{2}+\eta_{i}$. Since we are assuming a steady state growth, it is required that 
for the given time interval $\Delta t$, the distribution of $\Delta h_{i}$, which is 
same as that of $(\delta x_{i})^{2}+\eta_{i}$,  
must be independent of time. The time independence indicates that the random force 
as represented by the noise term is adequately compensated for by the stabilizing 
growth term. The time dependence for the distribution will indicate that on 
the growing surface 1) the 
weightages of configurations are changing in time, and/or 2)new configurations are 
generated as growth proceeds affecting the morphology on given scale.  
This will imply that the {\it true steady state} is not obtained. Normally one identifies 
steady state region by inspecting $w$ vs. $t$ on logarithmic scale. The beginning of 
linear region on this plot is considered as the onset of steady state region. 

In general,  $\Delta h_{i}$ is a result of all the terms on the r.h.s. of a growth 
equation. Hence, 
cross over regions will not be discriminated with respect to the distribution 
of  $\Delta h_{i}$.  The distribution of  $\Delta h_{i}$ at small and large length scales 
may differ. However, in either case it must be time independent irrespective of the 
dominant term at the given scale. 

 Local configuration defining growth term is directly related to 
 $\Delta h_{i}$. This suggests that a measure of  $\Delta h_{i}$ can be obtained 
by defining local height with respect to a local reference. Such height will respond to 
local changes in heights and help in providing a measure of  $\Delta h_{i}$.  
We define such a height as a height measured from average height of neighbors.   
$(h_{i})_{local}=h_{i}-(h_{i-1}+h_{i+1})/2$, proportional to the difference 
between the local steps. Incidentally, the expression is similar to the second derivative
of $h(x)$ representing EW term. However, for any growth term, same expression can be 
used in defining a measure of $\Delta h_{i}$. The definition $(h_{i})_{local}$ is more general. 
Note that $(h_{i})_{local}=(\delta x_{i}-\delta x_{i-1})/2$. 
 In 1+1- dimensions, the (h-h) correlations in the discrete form are 
\begin{equation}
G(n)=\left<(h_{i}-h_{i+n})^{2}\right>
\end{equation}
We assume that {\it the correlation length $\xi$ is very large compared to 
 the length $a$}. 
Using the definition of steps Eq.(4),  
\begin{equation}
G(2)=<\delta x_{i}^{2}+\delta x_{i+1}^{2}
+2\delta x_{i}\delta x_{i+1}>. 
 \end{equation}
Let $<\delta x_{i}^{2}>=<\delta x_{i+1}^{2}>=\delta^{2}$ and , $<\delta x_{i}
\delta x_{i-1}>=s\delta^{2}$ where $s$ is the coupling between the 
steps around a given site $i$. The distribution for $\delta x_{i}$ is {\it always}  
symmetric around zero and time independent for an 
ensemble average.
 
 In the limit $\xi \rightarrow \infty$  Eq. (3) reduces to 
 $G(x)=cx^{2\alpha}$ where constant $c=G(1)$. Hence Eq. (6) can be written 
as 
\begin{equation}
2^{2\alpha}=2+2s
\end{equation} 
where $G(1)=\delta^{2}$ in the discretized case. Coupling $s$ uniquely 
determines $\alpha$. Thus, for $s$=-1/2,0,and 1, $\alpha$ is 0, 0.5, and 1 
respectively. This analysis can be easily extended to higher dimensions. The 
relation between $\alpha$ and $s$ remains unchanged over a square, cubic or hypercube 
lattice in higher dimension. We have independently verified it for the EW model.  
A rough surface will be characterized by some value of $s$. 
Expressing $(h_{i})_{local}$ in terms of steps it can be shown that
  $4<(h_{i})_{local}^{2}>=(2-2s)\delta^{2}$ . 
This shows that the distribution of $(h_{i})_{local}$
can be used as a measure of $\alpha$ and the   
  definition of 
$(h_{i})_{local}$ is applicable to any growth equation which has $0\le \alpha \le 1$.    
However,  distribution of $(h_{i})_{local}$ cannot be used to represent that of  $\Delta h_{i}$.
Latter quantity is a result of local configuration dependent term {\it and} the fluctuation due 
to the noise term. Thus the appropriate measure of the  distribution of $\Delta h_{i}$ will 
be, uncorrelated fluctuations in $(h_{i})_{local}$. Hence, we measure the distribution of 
change in $(h_{i})_{local}=\delta (h_{i})_{local}$ over a time interval $\Delta t>w(t)$
, ensuring the uncorrelated fluctuations. 
 For any model expected to follow the KPZ , EW or any other conservative or non conservative 
growth equation, 
it is required that this distribution must be constant in time.

 In the following we apply this criterion of constancy to 
the same models described earlier. Although above discussion was for growth on one dimensional 
substrate , the corresponding criterion can be easily extended to higher dimensions.
We have chosen $\Delta t=$
100 MLs in  1+1 dimensions and 60 MLs in 2+1 dimensions where $\Delta t$ is the 
time difference considered in the simulations  . The time difference over which the 
constancy of distribution is tested is from 500 MLs to 5000 MLs in 1+1 as well as 
in 2+1 dimensions.   
In this time interval for all the models considered, 
$ln (w)$ Vs $ln (t)$ curve is linear implying a steady 
state growth region. In 2+1 dimensions, two sets of  $\delta (h_{i})_{local}$ are 
generated, one corresponding to $x$ and other $y$ direction, and added. The constancy of the 
distribution is checked by measuring the ratio of the values in the distribution 
 at zero for 500 MLs and 5000 MLs {\it i.e.} $P_{0}=100(\frac{I_{500}-I_{5000}}
{I_{500}})$, where  $I_{t}$ is the count of $\delta (h_{i})_{local}$ 
at zero. The ratios $P_{0}$ are obtained by averaging 
over large enough runs so that values of $P_{0}$ are statistically discriminated. 
In 1+1 dimensions over 3000 runs are required for substrate size of $L=8000$. We 
have also calculated sum of absolute values of $P_{i}=100(\frac{I_{i,500}-I_{i,5000}}
{I_{i,500}})$ measured between i=-4 to +4 
values of the $\delta (h_{i})_{local}$ as an additional measure of the constancy of the 
distribution of $\delta (h_{i})_{local}$. Here, $I_{i,500}$ and -$I_{i,5000}$ are the 
counts at $i$th position in the distribution at 500 and 5000 MLs respectively.  
This range (-4 to +4) is chosen because one of the models 
used in the present work provides values of $\delta (h_{i})_{local}$ only 
within this range. This sum is denoted by $P_{sum}$.

 In Table I we present the results from the measurement of distribution of 
$\delta (h_{i})_{local}$ for different models in 1+1 dimensions. Along with 
 $P_{0}$ and $P_{sum}$, we also measure $\alpha$ for each model. However, 
as is seen from Fig. 1, $G(x,t)$ for KK and NN1 models are curved on the 
log-log scale. Hence, the measured $\alpha$ values are not reliable.    
Fig.3 shows a plot of such a distribution for the model NN1 described above.

 KK Model: It shows a deviation of -0.5$\pm0.06$\% and 6.86$\pm1.0$\% in  $P_{0}$ and 
$P_{sum}$ respectively. For KK model the distribution becomes narrower in 
time (-ve $P_{0}$) indicating that 
number of tilted regions are growing with time. We have used $N=1$ for KK model as 
height limitation \cite{kk}.

SC Model:  With $\sigma=1.7$  it shows significantly small spread compared to KK 
model in the distribution. 
We have observed that for other values of  $\sigma$, spread is larger. 
The advantage of this model is that it does not suffer 
from cross over effects. By varying $\sigma$ it is possible to get faster convergence 
to the KPZ dynamics.

 We have performed simulations between 
200 MLs to 2000 MLs, 2000 MLs to 20000 MLs. Corresponding results have same trends 
{\it i.e.} for SC model the $P_{0}$ and $P_{sum}$ are less than 0.1\% and 1.5\%
respectively  
while for KK model the absolute value is larger than 0.5\% and 6\%. Our 
analysis is based on the spatial discretization 
of the growth equation. It is known that \cite{hh} the real space group renormalization
leads to different results than the KK result. Also for the KK model, in reference \cite{ala} 
the $\alpha$ value for 1+1- dimensions is 0.489 although in \cite{kk} from saturation 
of width at larger $L$ values it is given as 0.5. Our results show that the 
pathology associated with KK model is manifested in the form of change in the 
distribution of $\delta (h_{i})_{local}$ in time. This in turn indicates that the finite 
size effects are dominant over the time scales considered. 

The results for conservative 
growth models also confirm to this behavior.
 
HM Model and NN1 Model: HM model is like solving linear second 
ordered growth equation locally. It is expected to follow the dynamics exactly. 
The $P_{0}$ and $P_{sum}$ values are indeed close to zero for this model. The NN1 model 
restricts the minimization of $h(x)$ due to the constraint that it is immobile if one or 
more nearest neighbors are present. This model has larger values of  $P_{0}$ and $P_{sum}$.

Above comparison between HM and NN1 models show that 
 finite size effects do not allow convergence to EW dynamics in NN1 model due to 
the constraint over the time scale studied. We consider the 
deviation from zero for  $P_{0}$ and $P_{sum}$ as the measure 
of convergence to the universality class that the model is belonging to. Larger the 
deviation farther is the model from the convergence. This convergence is with respect to the 
true dynamics of the growth equation, belonging to a particular universality class. 
Thus, SC model converged to the KPZ universality within the statistical 
error bar for the measurements of  $P_{0}$ and $P_{sum}$, but KK model has not yet converged. 
Similarly, HM model has converged to EW universality but NN1 model has not. It is expected 
that asymptotically these convergences occur. However, within the time and length scales 
used for our measurements, scaling corrections do not allow true value measurements 
for KK and NN1 models in the KPZ and EW universality classes respectively.   
        
Above results assert that a 
model can be considered to have converged to its representative growth equation 
dynamics if the spread in its  
distribution of  $\delta (h_{i})_{local}$
is minimum. The advantage of this measurement is that it can be performed over a 
relatively small amount of growth compared to the $G(x,t)$ measurement. Thus, even in 
higher dimensions the method can be used to determine the finite size effects in a model 
at an early stage.   

We have applied this method to determine an accurate value of $\alpha$ for the KPZ equation 
in 2+1 dimensions. For KK (N=1) model the $P_{0}$ is 
$0.29\pm0.02\%$ and $P_{sum}=5.0 \pm 1.0\%$ with 
 $\alpha=0.402\pm0.016$.  
For SC model with $\sigma=2.5$, $P_{0}$  is $0.02\pm0.05\%$ and $P_{sum}=1.3 \pm 1.1\%$
with $\alpha=0.355\pm0.001$. For KK model $\alpha$ is measured between $x=5$ and 20 while 
for SC model it is between $x=2$ and 50.    
We have plotted these results in 
Fig. 4.   
These results show that in 2+1 dimensions, 
true value of  $\alpha$ for KPZ equation is close to 0.36.

In conclusion, we have established a new criterion that can be applied to check the 
finite size effects at an early stage of growth. When applied 
to the models representing KPZ equation, it is seen that a new model , 
SC model in this connection 
can be adjusted to follow the KPZ dynamics accurately. This is true in both 1+1 and 
2+1 dimensions. Based on this study it is seen that KK model predicts exponents 
 that are away from the true values. Only those models that satisfy the condition of 
invariance of the distribution of $\Delta h(x)$ will follow the dynamics of the 
representative growth equation correctly. These are the models that have already  
converged to the respective universality in terms of the underlying dynamics. We find 
that SC model converge to the universality much earlier. The KK model 
is not converged over the time and length scales used here. The HM model converges 
to EW universality but NN1 does not. Hence, the exponent measurements based on 
converged models are reliable.

{}    
\pagebreak  

\begin{figure}
\label{Fig. 1}
\caption{Plot of $G(x,t)$ vs. $x$ for KK model(N=1) (open circles), 
for HM  model (open squares) , for NN1 model (filled circles), and for SC model (filled squares)
in 1+1 dimensions. The curves are shifted along y-axis to avoid overlapping data. 
The growth is over 5.$10^{5}$ MLs with L=80000. 
}
\end{figure}

\begin{figure}
\label{Fig. 2}
\caption{Plot of $c_{0}$ the straight line intercept on $y$- axis , 
as a function of average $ x$. The plots are for the models in (1+1) dimensions, KK(N=1)
(open squares), SC (+), NN1 (x) and, HM (filled squares).  
For the sake of comparison, values at $\bar x=15$ are adjusted to the same 
value for all the models.}
\end{figure}

\begin{figure}
\label{Fig. 3} 
\caption{Plot of distribution of  $\delta (h_{i})_{local}$
for the model NN1 on semi log scale. The inner distribution (+) is for 
$t=500$MLs and the outer one (x) is for $t=5000$ MLs.
}
\end{figure}

\begin{figure}
\label{Fig. 4} 
\caption{Plot of $G(x,t)$ vs. $x$ for KK model(N=1) (open squares),
and for SC model (filled squares)
in 2+1 dimensions.  
}
\end{figure}

\begin{table*} 
\caption{$\alpha$ values as obtained from height-height correlations, $P_{0}$ the ratios 
of values , and sum of ratios $P_{sum}$ of $\delta (h_{i})_{local}$ at 
500 MLs and 5000 MLs for different models
in 1+1 dimensions.}
\begin{tabular}{lccc} 
Model/Parameter & $\alpha$ & $P_{0}$ in \%&  $P_{sum}$ in \%    \\
&& \\ 
KPZ and \\
EW Equation & 0.5 & 0.0 &0.0\\ 
KK(N=1)  & 0.5089 $\pm$ 0.012 &  -0.5 $\pm$ 0.06 & 6.8 $\pm$ 1.0   \\ 
SC ($\sigma=1.7$) & 0.5062 $\pm$ 0.0015 & 0.07 $\pm$ 0.02 & 1.1 $\pm$ 0.4 \\ 
NN1  & 0.514 $\pm$ 0.02 & 1.82 $\pm$ 0.12 & 14.3 $\pm$ 2.0 \\ 
HM  & 0.496 $\pm$ 0.002 & 0.06 $\pm$ 0.05 & 1.5 $\pm$ 1.2\\ 
\end{tabular}
\end{table*}


\begin{thebibliography}{}
\bibitem{other} * Other 
references are available from the review articles like  
references 1) and 2) above.
\end{thebibliography}
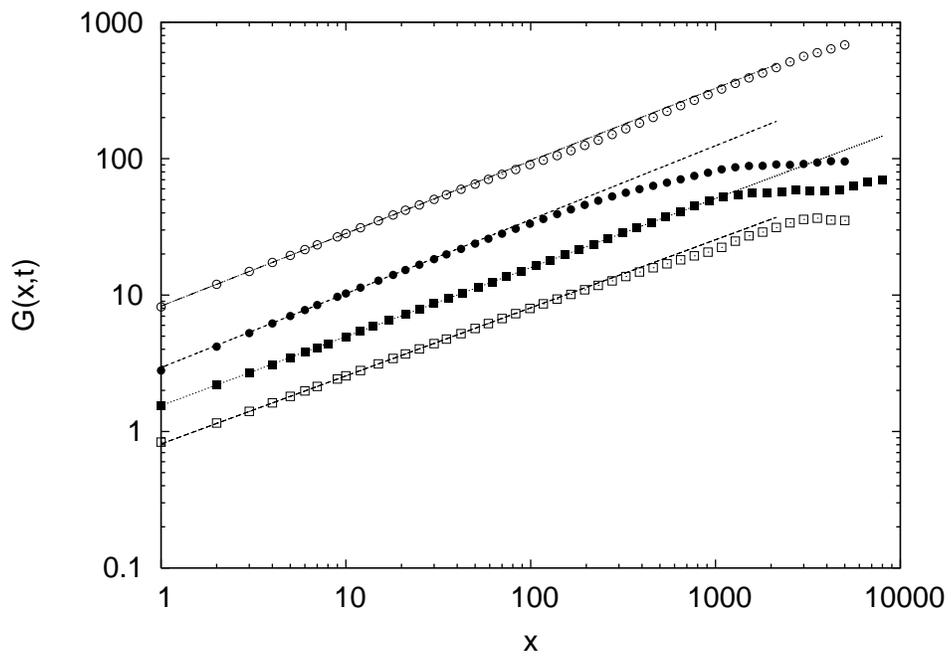
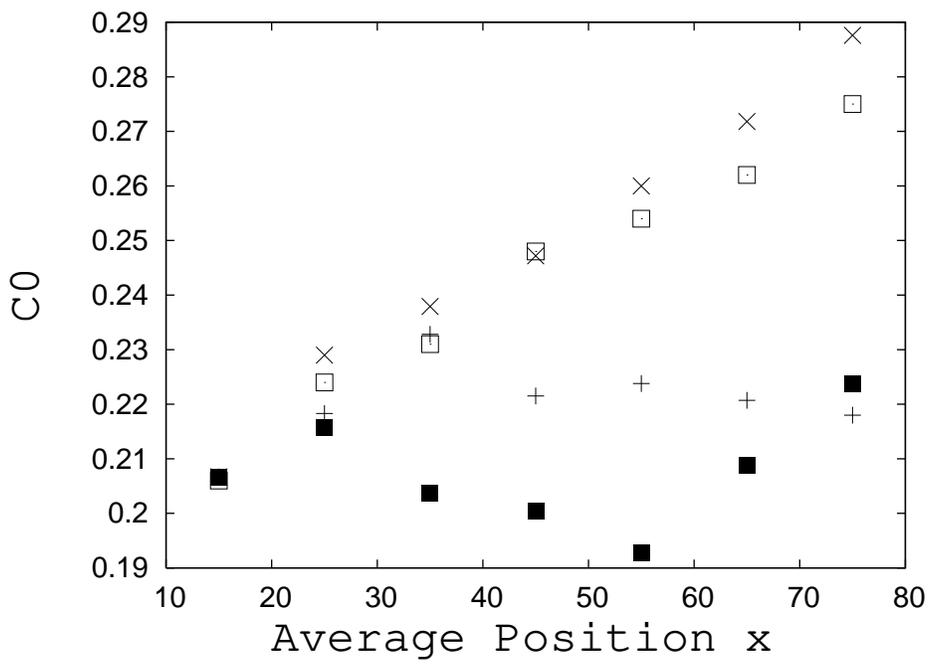
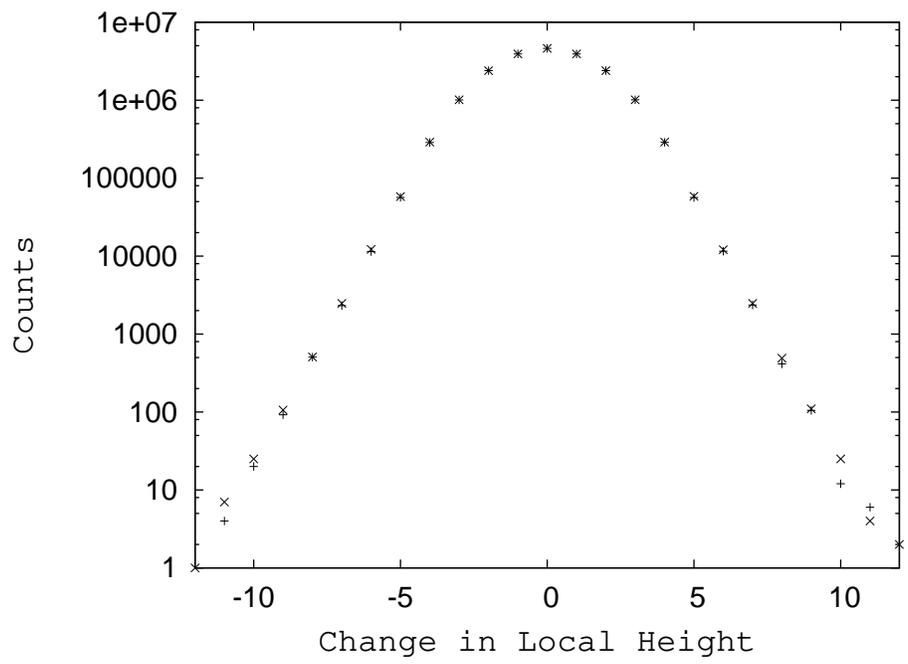
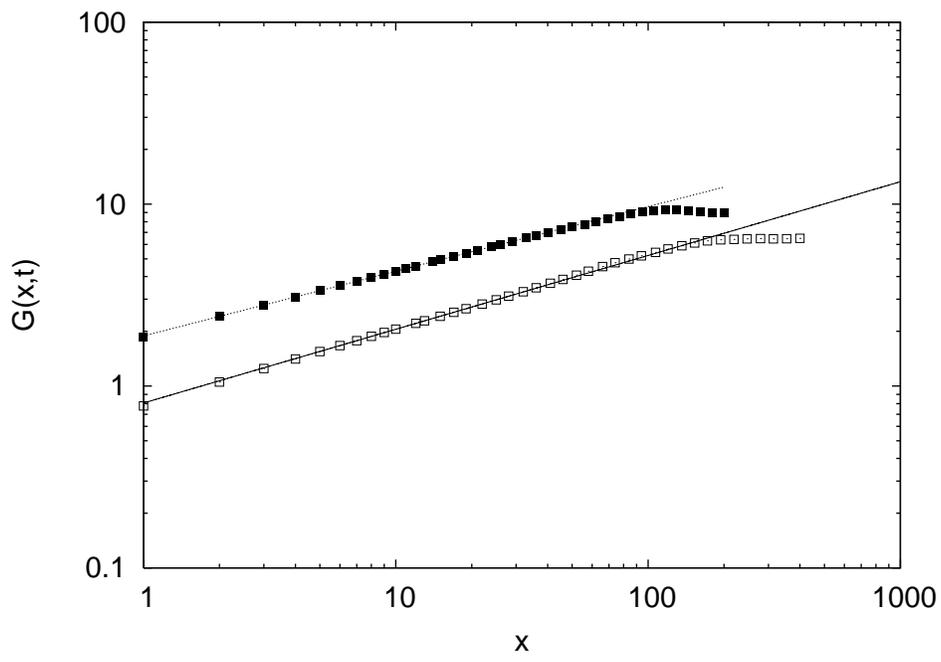
\end{document}